\documentclass{JHEP}
\input{epsf}
\usepackage{epsfig}

\newcommand{\opt}{\widetilde{O}^+}

\preprint{MIT-CTP-3163}
\title{Orientifold planes, affine algebras and magnetic monopoles}
\author{Amihay Hanany
\,
and Jan Troost
\\
\email{hanany@mit.edu,troost@mit.edu}\\ Center for Theoretical
Physics
\\ MIT
\\
   77 Mass Ave \\ Cambridge, MA 02139 \\
    USA\\
}
\abstract{We analyze string theory backgrounds that include
different kinds of orientifold planes and map out a natural
correspondence to (twisted) affine Kac-Moody algebras. The
low-energy description of specific BPS states in these backgrounds
leads to a construction of explicit twisted magnetic monopole
solutions on $R^3 \times S^1$. These backgrounds yield new
low-energy field theories with twisted boundary conditions and the
link with affine algebras yields a natural guess for the
superpotentials of the corresponding pure ${\cal N}=1$, and ${\cal
N}=1^{\ast}$ gauge theories.}

\begin{document}

\section{Introduction}
String theory backgrounds including orientifold planes have proven
to be a fruitful ground for study. Since their construction as
perturbative string backgrounds \cite{Sagnotti,Horava,DLP}, (see
\cite{Dabholkar} for a recent review on the perturbative aspects
of these objects) much has been learned on their non-perturbative
aspects. See \cite{HK} for a recent review, and further
developments can be found in \cite{BGK,BKS,7author}. New types of
orientifold planes are discovered by turning on various discrete
fluxes, changing the charges and tensions of these objects. By
employing a general principle of branes ending on branes, it is
possible to introduce ``brane realizations of Discrete Torsion''
\cite{HK}. This allows for a simple intuitive understanding of the
various types of orientifold planes present in the string theory
backgrounds. Especially interesting is the classification of
disconnected components on the moduli space of compactifications
for a small number of compact dimensions. For backgrounds with 16
supercharges this question received some answers
\cite{BKS,7author,Keurentjes,HJK}. Such backgrounds are partially
classified by the rank of the gauge group,\footnote{Rank reduction
was first discussed in \cite{BPS}.} alternatively by the number of
massless vector multiplets in the low energy theory. Let us review
some of these results. In 10 dimensions there are 2 theories, each
with rank 16 and with a single scalar in a gravity multiplet
(dilaton). One is the $Spin(32)$ theory and the other the
$E_8*E_8$. No continuous parameter connects these theories in 10
dimensions. In 9 dimensions there are 3 disconnected components,
with ranks 17, 9, 1. These, for example, correspond to M theory
compactifications on a cylinder, M\"obius strip and Klein bottle,
respectively. In 8 dimensions we have 3 components with ranks 18,
10, 2, etc. The analysis can be extended down to very low
dimensions as recently studied in \cite{Arjan}. The present paper,
however, will follow a different route.

As for some of the low rank cases above, we will be interested in
studying backgrounds including different types of orientifold
planes. The mere fact that $D$-branes placed in this background
have different enhanced gauge symmetries, depending on their
location relative to the different orientifold planes, leads to
complications in formulating the low-energy theory on these
$D$-branes. For example, in a background which contains an $O^+$
plane and an $O^-$ plane, the enhanced gauge group is $Sp$ near
the $O^+$ plane and is $SO$ near the $O^-$ plane. The gauge theory
on the D brane can be written, locally, at the vicinity of each of
the orientifold planes, but not in a global fashion which will
take both gauge enhancements into account. A clue to resolving
this issue comes from analyzing certain BPS-states that exist in
these backgrounds -- this lays bare a natural connection to affine
algebras (see e.g. \cite{Harvey:1998gc,DeWolfe:1999pr} and
references therein for earlier analysis of infinite algebras in
the context of string theory). In this paper we show that this
analysis leads to a stringy realization of (twisted) affine
algebras, and periodic instanton and monopole solutions in
low-energy theories with twisted boundary conditions.

The outline of the paper runs as follows. In section
\ref{reviews}, we review the guise in which monopoles appear in
$D$-brane settings including orientifold planes. In section
\ref{oplanes} we combine the building blocks from section
\ref{reviews} to make a connection between orientifold planes and
affine algebras. Here we present a correspondence between a set of
affine and twisted affine algebras and their corresponding
representations in terms of branes and orientifolds. While the
construction is very simple, it gives a very powerful realization
of infinite dimensional algebras on branes in string theory. In
particular, we point out that the introduction of a compact direction
transverse to a set of D branes turns the finite dimensional gauge
symmetry on the branes to an infinite dimensional one. In section
\ref{subalgebras} we show how the enhanced gauge symmetries at the
orientifold planes are encoded in the algebraic framework. We
study various limits on the moduli space and find agreement with
the string theory expectations. Section \ref{field} is devoted to
finding explicit descriptions of BPS states in the low-energy
field theory and section \ref{conclusion} contains conclusions. In
appendix \ref{app} we summarize our conventions.

\section{Review of $D$-branes and monopoles}

\label{bb} \label{reviews} In this section we review the relation
between BPS monopoles and $D$-branes in a simplified setting, and
introduce the building blocks that are needed in the next section
for a more detailed analysis of less well-known string theory
backgrounds.
\begin{figure}
\begin{center}
\epsfbox{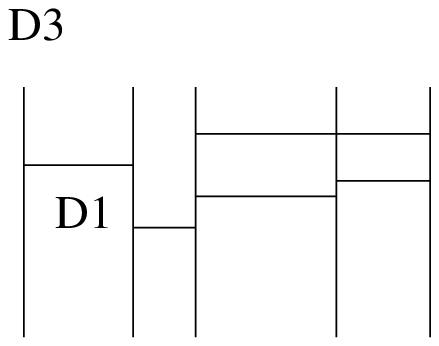}
\caption{$D3$-branes separated on a line with $D1$-branes
stretching between them.} \label{D3D1}
\end{center}
\end{figure}

A string vacuum consisting of $N$ parallel, separated and
co-linear $D3$-branes (figure \ref{D3D1}), preserving 16
supercharges, contains non-perturbative BPS states that correspond
to $D1$-branes stretching between $D3$-branes. In the low-energy
field theory on the $D3$-branes these correspond to magnetic
monopole solutions in a spontaneously broken gauge theory
\cite{Strominger:1996ac,Douglas:1996du,Diaconescu:1997rk}, where
the scalar vacuum expectation value at infinity sets the
separation between the $D3$-branes. The fundamental monopole
solutions \cite{Weinberg:1980zt} (in the background of which the
Dirac operator has only four zero modes) correspond in figure
\ref{D3D1} to $D1$-branes stretching between two neighboring
$D3$-branes. On the other hand, it is intuitively clear that a
$D1$-brane stretching between two $D3$-branes that are not
neighbors will have more zero modes since it can split into one or
more $D1$-branes (see the right hand side of figure \ref{D3D1}).
For such a case the corresponding monopole solution is not
fundamental. The fundamental monopole solutions can be labeled by
the simple co-roots of the gauge group \cite{Weinberg:1980zt}
 associated to the $D3$-branes. For our $SU(N)$
 example\footnote{We ignore the center of mass $U(1)$ for now.}, there are $N-1$
simple co-roots, in one to one correspondence with $D1$-branes
stretching between neighboring $D3$-branes. S-duality leads to
similar reasoning for the electrically charged macroscopic
$F1$-strings. These are associated with the roots of the gauge
group.
\begin{figure}
\begin{center}
\epsfbox{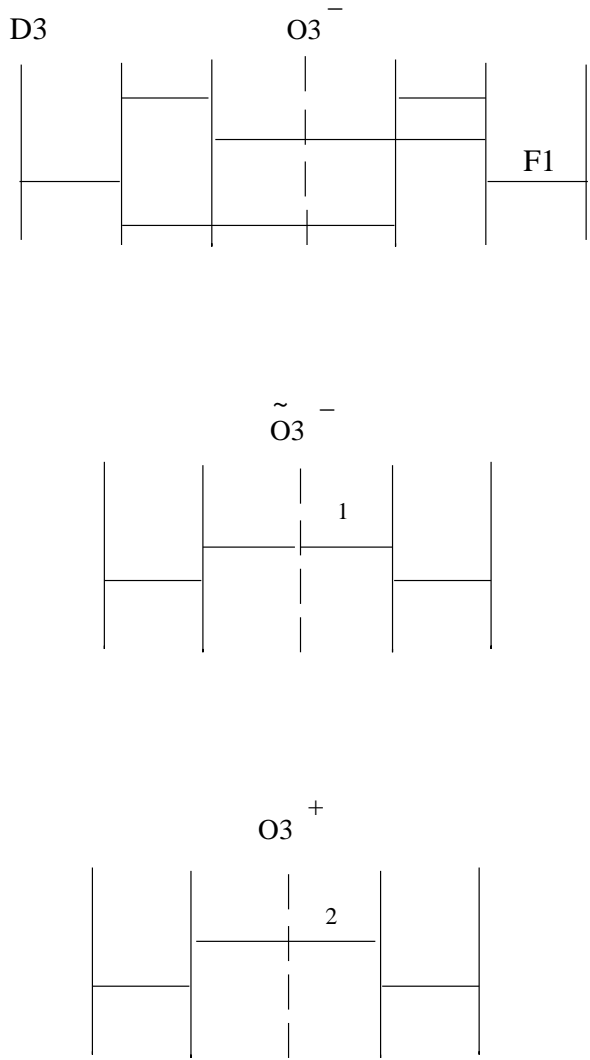}
 \caption{The three types of
orientifold planes and allowed fundamental $F1$-string BPS states
in the covering space. The relative length of the $F1$-strings in
the cases of the $\widetilde{O3}^-$ and $O3^+$ is indicated above
the $F1$-strings. Note that mirror images of the horizontal
strings represent the same state (and that only the string in the
middle of the lowest figure is its own mirror image). The
$\widetilde{O3}^+$ plane has the same electric spectrum as the
$O3^+$ plane. \label{coverblocks}}
\end{center}
\end{figure}

When we include orientifold planes (the orientifold planes which
will be considered are the $O^-$, $O^+$, $\widetilde{O}^-$ and
$\widetilde{O}^+$, as reviewed in \cite{HK}) in the string
background, the analysis changes slightly. In the presence of
orientifold planes, some states are projected out of the string
spectrum, and the analysis is repeated taking into account that
only certain $D1$-branes survive as BPS states in the spectrum. In
the S-dual picture, which admits a perturbative analysis and is
more convenient for our purposes, we can understand the absence of
certain states as due to the orientifold projection  condition
\cite{Pradisi:1989xd,Gimon:1996rq}
 that yields only a subset of invariant $F1$-string states.
In figure \ref{coverblocks} we depict the covering space and draw
some of the fundamental $F1$-string states. The difference between
the string passing the $O^+$ plane and the string ending on the
$\widetilde{O}^-$ plane is indicated by their relative length.
Note, for example, that an $F1$-string can end on an
$\widetilde{O}^-$ plane, since there is a $D$-brane stuck on that
orientifold plane, while it can not end on an $O^+$ plane.

\begin{figure}
\begin{center}
\epsfbox{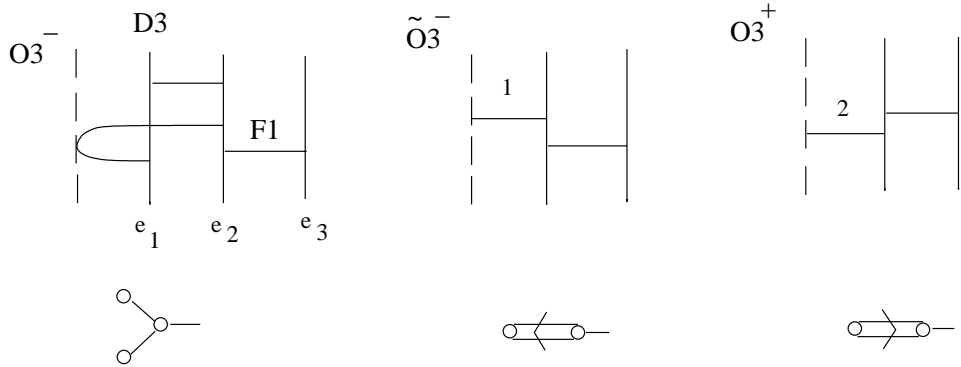}
\caption{The building blocks for
Dynkin diagrams and their corresponding orientifold planes. We
indicated the positions of the $D3$-branes in the $O3^-$-plane
example.} \label{buildingblocks}
\end{center}
\end{figure}

The simple roots corresponding to the fundamental electrically
charged solutions \cite{Weinberg:1980zt}, can be represented by
the distances between the $D$-branes on which they end. Their
natural intersection form is represented by a Euclidean
intersection form for the positions of these $D$-branes. For
example, we can consider the fundamental $F1$-strings in figure
\ref{buildingblocks} (where we draw the space resulting after the
orientifold projection) for the $O3^-$ case. They can be
associated to the positions of the $D3$-branes on which they end.
We find the vectors: $e_3-e_2, e_2-e_1$ and $e_2+e_1$, which have
intersections proportional to the intersections of the simple
roots corresponding to the nodes drawn in figure
\ref{buildingblocks}.

We can summarize the correspondence between orientifold planes
combined with $D$-branes, and the simple root system by
associating (part of) a Dynkin diagram to each orientifold plane
as in figure \ref{buildingblocks}. The root system is associated
to the electrically charged states in the corresponding low-energy
gauge theory. The electrically charged BPS states in the case of
the $\widetilde{O}^+$ plane span the same root system as for the
$O^+$ plane. We return to the difference between these two cases
in the next section. It resides in the magnetically charged states
that correspond to a co-root system. The example we gave in this
section in terms of $O3$-planes, $D3$- and $D1$-branes can be
T-dualized to other $O(p+2)$-planes, $D(p+2)$ and $Dp$ systems for
$p\le6$, in the cases where such objects exist.
\section{Orientifold planes and affine algebras}
\label{oplanes} In the previous section we reviewed the building
blocks that are put to good use in the present section. We combine
the blocks by studying string theory backgrounds including
parallel orientifold $p$-planes and $Dp$-branes. The background
geometries we study include an interval with two orientifold
planes at the endpoints, each can be of type $O^-, O^+,
\widetilde{O}^-$ or $\widetilde{O}^+$. In between the orientifold
planes we place $Dp$-branes, aligned with the two orientifold
planes. In the covering space we have two orientifold planes on
the fixed points of a circle under the orientifold action together
with D-branes and their mirrors.

\begin{figure}
\begin{center}
\epsfbox{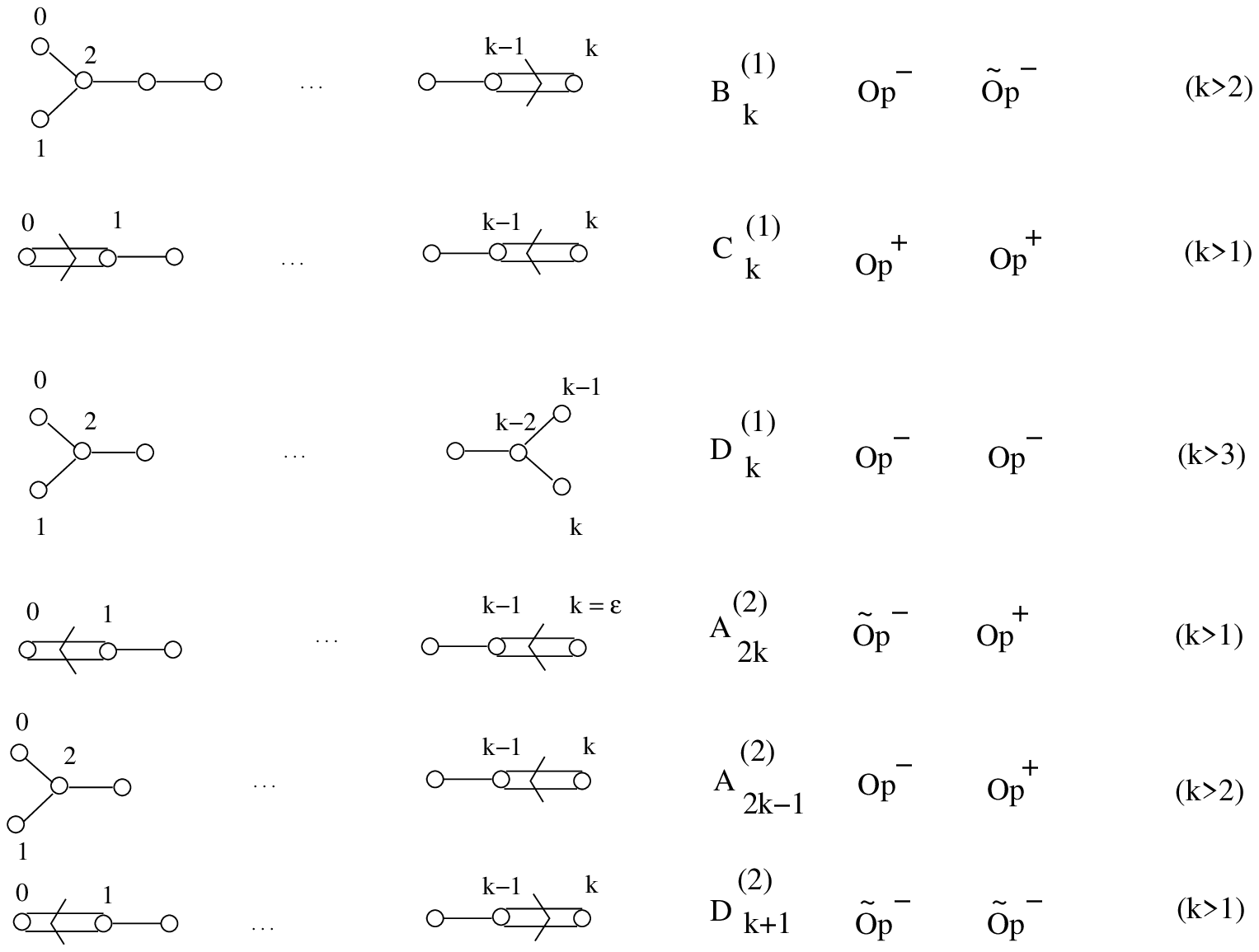}
\caption{On the far left we denote the Dynkin diagrams of
untwisted and twisted affine Kac-Moody algebras. The first 3
diagrams represent affine algebras while the last 3 represent
twisted affine algebras. Their standard notation is next to the
diagrams, where the superscript in the notation for the (twisted)
affine algebras denotes the order of the outer automorphism that
is used to construct them \cite{K} (see figure \ref{outerauto}).
The six different combinations of the orientifold planes $Op^-$,
$\widetilde{Op}^-$ and $Op^+$ used to construct the Dynkin
diagrams within string theory, is shown in the next column. On the
far right are limitations on the rank of the affine algebras for
the Dynkin diagram to make sense. The zeroth root always
corresponds to the root $\alpha_{\epsilon}$ to be defined later,
except for the case $A_{2k}^{(2)}$.} \label{orienalge}
\end{center}
\end{figure}

\begin{figure}
\begin{center}
\epsfbox{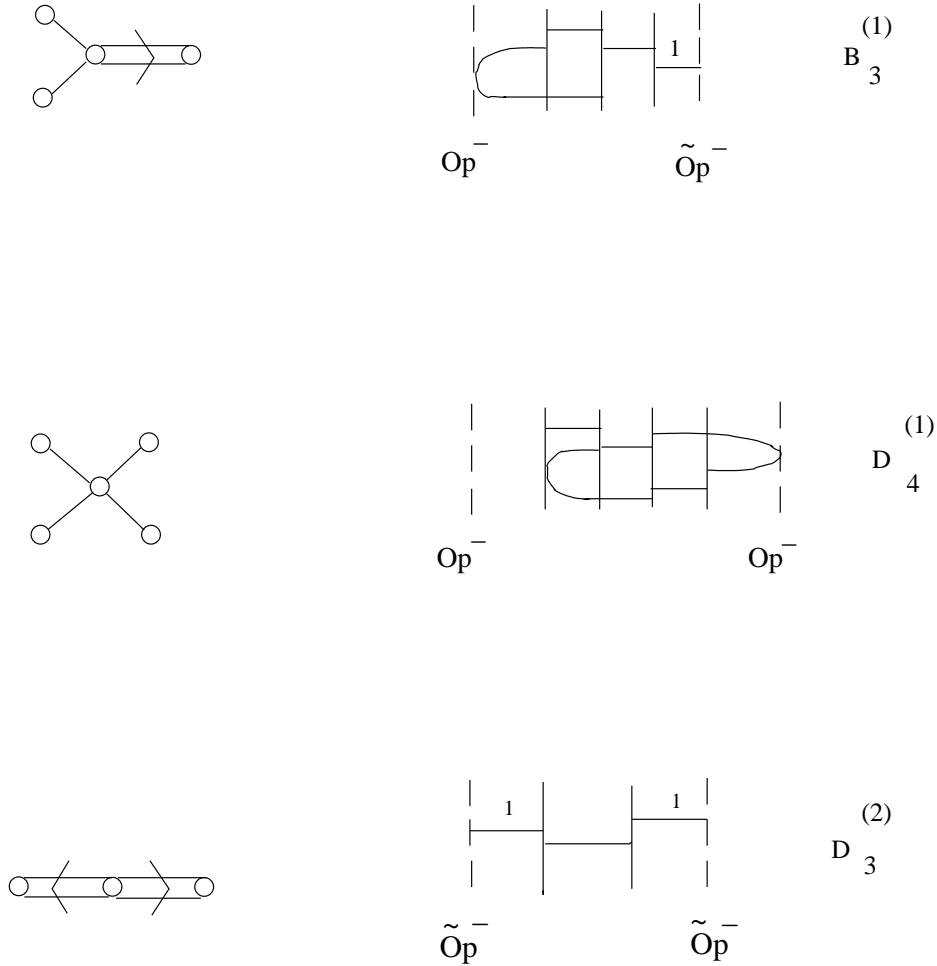}
 \caption{A
few examples of combinations of orientifold planes with a minimum
number ($l$ of figure \ref{orienalge}) of parallel D-branes
between them.} \label{exampleorien}
\end{center}
\end{figure}

Using the building blocks introduced in section \ref{bb} and their
magnetic associates, we can already press home one of the main
points of our paper. First of all, concentrating on the
electrically charged states, we can construct the Dynkin diagrams
corresponding to the simple roots associated to the $F1$ branes
stretching between the $Dp$-branes and the orientifold planes. We
draw these Dynkin diagrams in figure
\ref{orienalge}.\footnote{Some low-rank examples are shown in
figure \ref{exampleorien}.} We recognize these Dynkin diagrams as
corresponding to six infinite series of (twisted) affine algebras
\cite{K} (see also \cite{KM,Goddard:1986bp,Fuchs:1992nq}). This
allows us to put forth a correspondence between string theory
backgrounds including two orientifold planes at the ends of an
interval, and (twisted) affine algebras, as displayed in figure
\ref{orienalge}. We will make the nature of the correspondence
more precise as we go along.

If in addition to electrically charged objects we consider
magnetically charged BPS objects, we can refine our correspondence
to make a distinction between the $O^+$ plane and the $\opt$
plane. The difference between $O^+$ and $\opt$ arises in their
monopole spectrum, i.e. in the magnetically charged
objects.\footnote{Restricting to $p=3$, under S-duality the $\opt$
plane is invariant. The complexified gauge coupling and theta
angle undergo the change $\tau\rightarrow\tau+1$, when the
discrete flux that shifts $O^+$ to $\opt$ is turned on.} Similarly
as in figure \ref{buildingblocks} (where the case of the $O^+$ and
$\opt$ coincide), we can draw a Dynkin diagram associated to the
magnetically charged states. Then the (ending of the) Dynkin
diagram of the $O^+$ plane and the $\widetilde{O}^-$ plane get
interchanged, but the $\opt$ plane remains associated with a
$C$-type ending.
 Making use of this reasoning based on S-duality (and of an obvious notation
indicating the three types of Dynkin diagram endings), we can summarize the
correspondences between the ten possible combinations of orientifold planes
and electric and magnetic charges in table \ref{ten}.
\begin{table}
\begin{center}
\begin{tabular}{|c||c|c|}
\hline \mbox{planes} & \mbox{electric} & \mbox{magnetic}
\\ \hline $--$ & DD & DD
\\ \hline $++$ & CC & BB
\\ \hline $\tilde{-} \tilde{-}$ & BB & CC
 \\ \hline$\tilde{+} \tilde{+}$ & CC & CC
 \\ \hline$-+$ & DC & DB
 \\ \hline$- \tilde{-}$ & DB & DC
 \\ \hline$- \tilde{+}$ & DC & DC
 \\ \hline$+ \tilde{-}$ & CB & BC
 \\ \hline$+ \tilde{+}$ & CC & BC
 \\ \hline$\tilde{-} \tilde{+}$ & BC & CC
 \\
  \hline
\end{tabular}
\caption{ On the left of the table we write the type of
orientifold $p$-planes at the two ends of the interval, in the two
following columns the algebras associated to the electrically and
magnetically charged BPS states. The letters indicate a $B,C$ or
$D$ type ending in the Dynkin diagram. For example, $DB$ stands
for the $B_k^{(1)}$ algebra. \label{ten}}
\end{center}
\end{table}

The appearance of affine algebras in the context of
compactification on a circle or an interval is no surprise.
Returning to our previous basic example of $D3$-branes (without
orientifold planes) with $D1$-branes stretched between them, we
notice that for $D3$-branes on a circle, there appears a new
fundamental magnetic monopole that stretches between the last
$D3$-brane and the first, completing the circle. It is well known
\cite{Garland:1988bv} that we can associate the extra fundamental
monopole to the extra root of  the affine algebra $A_{N-1}^{(1)}$,
thereby closing  up the Dynkin diagram of that algebra to form a
necklace. Notice that we have hereby completed a realization of
the infinite series of affine algebras, consisting of
$A_{k}^{(1)}, B_{k}^{(1)}, C_{k}^{(1)}, D_{k}^{(1)}, A_{2k}^{(2)},
A_{2k-1}^{(2)}, D_{k}^{(2)}$ \cite{K} in string theory
backgrounds. Apart from these, there are a few special low-rank
and exceptional cases that we will briefly return to.

\section{Gauge enhancements and subalgebras}
\label{subalgebras}

In this section we initiate a more detailed description of the link between
the orientifold backgrounds and the (twisted) affine algebras. In the
following, we concentrate on the electrically charged states. We will also
make good use of the T-dual of these configurations along the interval
separating the orientifold planes. After T duality a pair of orientifold
planes combines into a single object which can be either an ordinary
orientifold or a more exotic object. The results of T-duality are summarized
in table \ref{cor}.

\begin{table}
\begin{center}
\begin{tabular}{|c||c|c|c|}
\hline & $O^-$ & $O^+$ & $\widetilde{O}^-$ \\ \hline \hline $O^-$
& $O^- [D_k^{(1)}]$ & $o [A_{2k-1}^{(2)}]$ & $\widetilde{O}^-
[B_{k}^{(1)}]$ \\ \hline
 $O^+$ & $o [A_{2k-1}^{(2)}]$ &
$O^+ [C_{k}^{(1)}]$ & $\widetilde{o} [A_{2k}^{(2)}$] \\ \hline
$\widetilde{O}^-$ & $\widetilde{O}^- [B_{k}^{(1)}]$ &
$\widetilde{o} [A_{2k}^{(2)}]$ & $O^- + \mbox{a}\,
$D$\mbox{-brane} [D_{k+1}^{(2)}]$ \\ \hline
\end{tabular}
\caption{``Addition table'' for orientifolds. On the top (left) of the table
we write the type of orientifold $p$-plane which is placed at the first
(second) end of the interval. They combine after T-duality along the interval
to an orientifold $p+1$-plane that is written in the body of the table. We
also include the correspondence to affine Kac-Moody algebras in square
brackets. \label{cor}}
\end{center}
\end{table}
The notations $\widetilde{O}^-, O^+, O^-$ indicate familiar
orientifold $p+1$ backgrounds, and $o, \widetilde{o}$ indicate
backgrounds including modding out by a shift (more detailed in
\cite{DP,Witten:1998bs,Keurentjes:2000bs,Gimon,Hori}).
Concentrating on the algebra associated to the electric charges,
we include the correspondence between the orientifold planes and
the affine Kac-Moody algebras (see table \ref{cor} and figure
\ref{orienalge}). Since the occurrence of twisted affine algebras
is new, we discuss especially the properties of the string
backgrounds corresponding to the twisted affine algebras in more
detail in the following. The combinations of orientifold planes
for which the T-dual corresponds to a well-known string theory
background are $(O^-,O^-)$, $(O^+, O^+)$ and
$(O^-,\widetilde{O}^-)$. We do not treat these cases in detail.
One can make a similar but more straightforward analysis,
analogous to the cases discussed in detail in the next sections.

\subsection{$(O^+,O^-)$ and $(O^+, \widetilde{O}^-)$ orientifolds}

In this subsection we analyze how the finite subalgebras of the
affine algebras surface as enhanced gauge symmetries in the string
backgrounds. We show how they arise naturally from the vector
modes in the open string sector when we assume the action of an
outer automorphism on the Chan-Paton matrices of the open
strings.\footnote{This is reminiscent of the construction of CHL
strings in \cite{Chaudhuri:1995fk,Chaudhuri:1995bf} .} We show
that this assumption is consistent with our intuition for enhanced
gauge symmetries near orientifold fixed planes.

First, we discuss in detail the case of the $(O^+,O^-)$ and $(O^+,
\widetilde{O}^-)$ orientifolds. We study the T-dual of these
configurations, which correspond to $o$- and $\tilde{o}$-planes
(see table \ref{cor}). The orientifold action includes in this
case a shift along the interval. We concentrate on the open string
modes in these backgrounds since we do not expect new issues to
arise in the closed string sector. The bulk closed string theories
for these backgrounds are the usual Type II theories (for $p\le
9$).

\begin{figure}
\begin{center}
\epsfbox{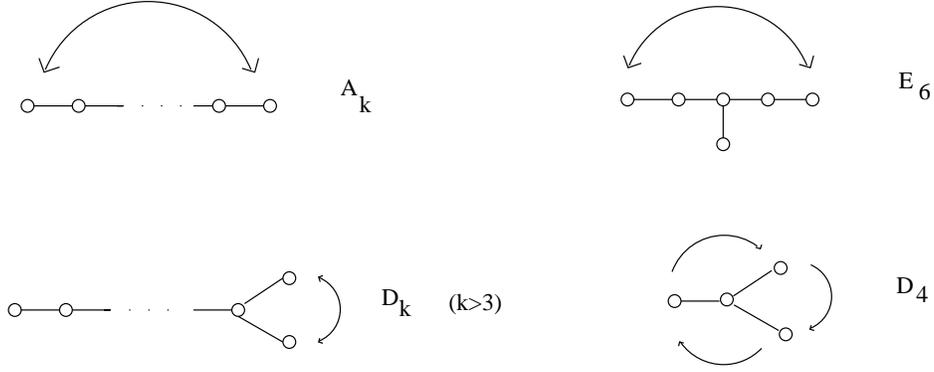}
\caption{ The
action of outer automorphisms on the simple roots of Lie algebras,
pictured here as an action on the Dynkin diagram, from which the
action on the full algebra follows. We also displayed two special
low-rank cases on the right.} \label{outerauto}
\end{center}
\end{figure}

In the open string sector, we first concentrate on the massless vector
states in the NS sector,
associated to the $D(p+1)$-branes coinciding with the $o$-plane.
We mod out the vector states in the covering theory
by $S$, the shift over $\pi R$ (where $R$ is the radius of
the covering circle), combined with $\Omega$ (orientation reversal
on the worldsheet),  but also with the action of
an outer automorphism $\tau$ on the Chan-Paton factors, as follows,
\begin{eqnarray}
\psi^{\mu}_{-1/2} |x\rangle_a T^a =+ \psi^{\mu}_{-1/2} |x+ \pi
R\rangle_a \tau(T^a) \label{oo}
\end{eqnarray}
where $\psi$ is the worldsheet fermion in the NS sector, the outer
automorphism acts by reflecting the $A_{N-1}$ Dynkin diagram, and where $N$
is the number of $D$-branes in the covering theory (which is even for the
$(O^+,O^-)$ case and odd for the $(O^+,\widetilde{O}^-)$ case). Figure
\ref{outerauto} shows the action of the outer automorphism on the Dynkin
diagram.
 The index $\mu$ is along the $D(p+1)$
worldvolume and  we
 have $\tau^2=1$.
This is our proposal for the action of the orientifold operation on the open
string states in the presence of an $o$-orientifold, and we will show it to
be consistent with intuitive expectations.

To describe the open string vector states after modding out, we define the
following subalgebras of the parent $A_{N-1}$ Lie algebra. We distinguish
the $g_{0}$-subalgebra (with elements $T^a_0$)
 corresponding to linear combinations of
generators in the Lie algebra with eigenvalue $1$ under the outer
automorphism, and the $g_{1}$-subalgebra (with elements
$T^a_1$)\footnote{The notation for the subalgebras is standard
\cite{K,KM}.} corresponding to generators with eigenvalue $-1$.
After modding out,
 we find the following
 open string states on the worldvolume of the $D(p+1)$-branes in the
 presence of
the $o$-orientifold:
\begin{eqnarray}
\psi^{\mu}_{-1/2} |2k\rangle_a T^a_0 \quad \mbox{and} \quad
 \psi^{\mu}_{-1/2} |2k+1\rangle_a T^a_1,
\end{eqnarray}
where $k$ is an integer specifying the momentum of the open string
state\footnote{All momenta are integer -- we define them
relative to the covering circle of length $2 \pi R$. The interval has
length $\pi R$.}. The parity of the momentum is paired with
the eigenvalue under the outer automorphism to insure invariance
under the combined orientifold operation (\ref{oo}).

 Note that we have not yet turned on any Wilson lines in the compact
direction on the worldvolume of the $D(p+1)$-branes. The massless
vector states on the worldvolume of the branes correspond to the
modes with zero momentum. They generate the $g_{0}$-subalgebra of
the twisted affine algebra $A_{N-1}^{(2)}$.
 These $g_{0}$-subalgebras are naturally associated
in the T-dual picture to D-branes located at a specific
orientifold fixed plane. Indeed, distinguishing the two cases
$A_{2k-1}^{(2)}$ and $A_{2k}^{(2)}$ we find the
$g_{0}$-subalgebras are $C_k$ and $B_k$ respectively \cite{K,KM},
indicating that the $Dp$-branes are located at the $Op^+$ plane
and $\widetilde{O}p^-$ plane respectively, when no Wilson lines
are turned on in the T-dual $o (p+1)$ picture.

We can now turn on Wilson lines. An interesting configuration
arises when we turn on the Wilson lines which correspond in the T-dual picture
to putting the D-branes at the position of the other orientifold $p$-plane.
  These Wilson
lines are (denoting the compactified direction $4$, and the gauge
field corresponding to the massless vector field in this direction
by $A_4$):
\begin{eqnarray} A_4&=&\check{\alpha_1} + 2
\check{\alpha_2} + \dots + (k-1) \check{ \alpha_{k-1}} + k
\check{\alpha_k}, \\ \nonumber A_4&=&\check{\alpha_{k-1}} + 2
\check{\alpha_{k-2}} + \dots + (k-i) \check{\alpha_{i}} +\dots +
(k-1) \check{\alpha_1} +\frac{k}{2} \check{\alpha_0},
\end{eqnarray} respectively\footnote{The numbering of the nodes in
figure \ref{orienalge} for the $g_{0}$-subalgebras agrees with the
standard
 conventions of \cite{OV}, except for the $A_{2k}^{(2)}$ case. This is natural
from the viewpoint of affine algebras since for all algebras but this one,
the $g_{0}$ algebra is obtained by deleting the zeroth node of the Dynkin
diagram (yielding the algebra known as $g^o$) \cite{K}. \label{ghato}}
\footnote{One way to find the appropriate
 Wilson lines is by looking
 at the positions of the D-branes and their mirrors in the covering
theory, and then follow these Wilson lines through the
construction of the twisted affine algebras \cite{K,KM}.}. These
Wilson lines give rise to a mass for one of the open string states
corresponding to the simple root $\alpha_k$, respectively
$\alpha_1$, but not to the states corresponding to the other simple
roots.
 Indeed, the states corresponding to the other simple roots are
not charged under the $U(1)$ associated to the Wilson line, as can
be seen by calculating their intersection with the Wilson line
$A_4$ (for example $\langle\alpha_2,A_4\rangle=0$), but the simple
root $\alpha_k$ (respectively $\alpha_0$) is. Moreover, there are
anti-periodic states which now become periodic (because of their
charge under the $U(1)$ associated to the Wilson line). They give
rise to a re-enlarging of the symmetry group that is appropriate
for the orientifold plane at which the $Dp$-branes are positioned.
Among these new extra massless states is the simple root
corresponding to the root that is needed to complete the $g_{0}$
algebra to the full twisted affine algebra. These read in the two
cases that we distinguish: $\theta_{hs}$ and $2 \theta_{hs}$ where
$\theta_{hs}$ is the highest short root of the $g_{0}$-subalgebra.
It is then straightforward
 to check (using the Cartan matrices of the $g_{0}$-subalgebras)
that indeed the periodicity of the state associated
to these roots changes from anti-periodic to periodic, by computing their
charge under the $U(1)$ associated to the Wilson line.
It is not difficult to infer the fate of the massless vector
states corresponding
to non-simple roots from the analysis for the massless vector states
 associated
to the simple roots.\footnote{It is amusing that for the $A_{2k}^{(2)}$ case,
there are now string modes with half-integer momentum, or T-dual winding, as
expected from the fact that strings can end on the $\widetilde{Op}^-$-plane.}
It is moreover easy to see how our analysis generalizes to other positions of
the $Dp$-branes on the interval.

It would be interesting to have a better understanding of the
geometric action of the outer automorphism. It is clear that it
acts as an orientation reversal in spacetime, but it would be nice
to have an interpretation of the associated signs
 (see e.g.\cite{Goddard:1986bp})
 in terms
of the background fields in string theory.

\subsection{$(\widetilde{O}^-,\widetilde{O}^-) $ orientifold planes}
The twisted algebra $D_{k+1}^{(2)}$ arises differently. From the
viewpoint of the parent theory, the Wilson line corresponding to
positioning all the D-branes at one of the
$\widetilde{O}^-$-planes, say the one at the origin, reads:
\begin{eqnarray}
\mbox{diag}(-\pi R,0,0, \dots, 0, \dots, 0), \label{matrix}
\end{eqnarray}
where in the parent theory we have $1+l$ D-branes corresponding to
the $2$ ${1\over 2}$ $D$-branes stuck to the orientifold planes
(in positions $1$ and $k+2$ in the diagonal matrix in
(\ref{matrix}))
 and $l$ branes and their mirrors. From the perspective of the parent
theory, we can see that the massless states corresponding to this positioning
of $Dp$-branes, and the positioning corresponding to the T-dual Wilson line
\begin{eqnarray}
\label{wl}
\mbox{diag}(-\pi R,-\pi R,\dots,-\pi R,  0, \pi R, \pi R, \dots, \pi R)
\end{eqnarray}
gives rise to the expected symmetry enhancements. These can be
reproduced by considering massless open string vector states as in
the previous section, but corresponding to the $D_{k+1}^{(2)}$
algebra. By turning on Wilson lines purely within this algebra one
cannot reproduce the same analysis as in the previous
subsection\footnote{The basic reason why the algebraic
interpretation of the parent theory is a bit less straightforward
in this case, is because of the non-trivial role of the center of
mass $U(1)$ which is not incorporated in a standard fashion in the
algebraic framework.}, but the embedding of the states in the
parent theory remains clear, and we can analyze the charges of the
states in the $D_{k+1}^{(2)}$ algebra under the Wilson line
(\ref{wl})  that includes a $U(1)$ piece. Then the analysis runs
as before. In this case the $g_{0}$-subalgebra is $B_k$, and the
algebra obtained after deleting the $k$-th node and adding in the
zeroth node is $B_k$, once again as expected in a
$(\widetilde{O}^-, \widetilde{O}^-)$ background. (Another way to
analyze this case is by copying the first part of the analysis
done for $A_{2k}^{(2)}$ and invoking the $Z_2$ symmetry of the
background.)
\subsection{Remarks}
We showed the connection between the infinite algebras and the enhanced
gauge symmetries that arise in the orientifold backgrounds. Before we
delve into some details of more explicitly exhibiting the solutions in
the low-energy theories corresponding to $D(p-2)$-branes spanning
between $Dp$-branes, we add some general remarks.

\begin{figure}
\begin{center}
\epsfbox{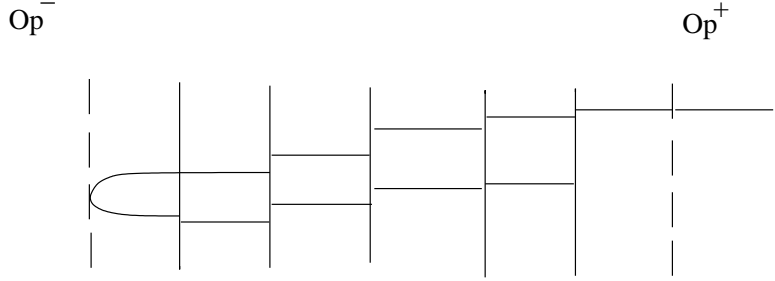}
\caption{We draw each $F1$-string with a multiplicity given by the
corresponding Kac label for the example $A_{2k-1}^{(2)}$ given in
the text. We replaced the label `2' over the $F1$-string near the
$O^+$ plane with an equivalent picture. The fundamental
$F1$-strings with these multiplicities make up a string that wraps
the covering circle.} \label{around}
\end{center}
\end{figure}

First of all, since 16 supercharges are preserved, the vector state we
discussed before should be thought of as being completed with fermionic and
bosonic states to form a full supermultiplet, and all other states will
similarly fall in supersymmetry multiplets. Next, when the interval has a
finite width, not all of the vector states can become massless at the same
time. The vector states we discussed will form an algebra which does not
yield a symmetry group, but an infinite spectrum generating algebra, which
should enable us to classify open string states at different mass levels into
representations of the affine algebras.
 We have already one example in the form of the $1/2$ BPS
states in this background, associated to the $D(p-2)$ branes
stretching between the $Dp$-branes. They are in one-to-one
correspondence with the roots of the infinite algebra, where the
Lie algebraic part indicates the brane they start from and end on,
and where the imaginary part indicates their winding number (or
T-dual momentum). The imaginary unit root, for example,
corresponds to the $D1$-brane fully wrapping the
circle\footnote{Except for the $D^{(2)}_{k+1}$ case, where it
spans the interval. Note that a string can start and end on the
$Dp$-branes stuck to the orientifold $\widetilde{O}^-$ planes at
the end of the intervals in this case.},
 as can be seen
from the fact that it corresponds to the sum of
all simple roots,
weighted with the Kac marks
of the affine algebra.
(For example, in the $A_{2k-1}^{(2)}$ case, we have Kac marks \cite{K}
$(1,1,2,2,\dots,2,1)$ and if we glue together fundamental $F1$-strings
with these multiplicities, we find that they correspond to a
$F1$-string going all the way around the circle (see figure \ref{around}).
This gives a nice pictorial interpretation to the Kac marks of affine
algebras.)
Using the vertex operators associated to the open string vector states, it
should be possible to build a full affine multiplet on any perturbative open
string state.

We also remark that there are more affine algebras than string
theory backgrounds that we discussed. Some special low-rank cases
correspond to a small number of $D$-branes between orientifold
planes, and we can find the correspondence to affine algebras
analogously to what we did before. We find for instance:
$D_3^{(1)}=A_3^{(1)}$, $D_2^{(1)}=A_1^{(1)}$ (as in the work
\cite{Kapustin:1998fa}), and $C_1^{(1)}=A_1^{(1)}=D_2^{(2)}$, and
the $k=1$ case of $A_{2k}^{(2)}$ indeed corresponds to the algebra
$A_{2}^{(2)}$. But the exceptional cases and the $D^{(3)}_4$ cases
are left without a string theory picture. It is tempting to
speculate that the last $D^{(3)}_4$ could arise from a $Z_3$
action on a covering string theory. Constructing backgrounds
corresponding to these algebras is an open problem.

We now turn to constructing explicitly some of the $D(p-1)$ branes
as generalized fundamental magnetic monopole solutions in the
low-energy field theory on the $D(p+1)$-branes.

\section{Field theory}
\label{field} We reviewed in section \ref{reviews} how $D1$-branes
stretching between $D3$-branes can be viewed as monopole solutions
in the $D3$-brane worldvolume gauge theory. After compactification
and T-dualizing along the direction in which the $D3$-branes are
separated (the $4$ direction), these solutions become solutions to
the self-duality equations for the field strength on $D4$-branes
spanning $R^3 \times S^1$. Note also that after T-duality a
$D1$-brane that originally wound the whole circle becomes a
$D0$-brane embedded in $D4$-branes, which in the low-energy theory
corresponds to an instanton in the gauge theory on the $D4$-brane
\cite{Douglas:1995bn}. We conclude
\cite{Garland:1988bv,Nahm:jb,Gross:1981br,
Lee:1997vp,Kraan:1998kp,Lee:1998vu}
that periodic instantons are made up of fundamental magnetic
monopoles. The low-energy descriptions for the fundamental
monopoles have been explicitly constructed in \cite{Lee:1998vu}
for all non-twisted affine groups and we will generalize this
construction to the twisted groups. The existence of these
solutions was mentioned in the mathematical treatment
\cite{Garland:1988bv}.

In section \ref{self-duality} we review the well known magnetic
monopole solutions corresponding to $su(2)$ subalgebras
\cite{'tHooft:1974qc,Polyakov:1974ek,Prasad:1975kr,Bais:1978fw}.
In section \ref{newsolutions} we show how to find new explicit
solutions to the self-duality equation on $R^3 \times S^1$ with
twisted boundary conditions, generalizing and simplifying
\cite{Lee:1997vp} \cite{Lee:1998vu}. In section
\ref{properties} we discuss properties of these new solutions that
identify them as fundamental monopoles
\cite{Weinberg:1980zt,Lee:1998vu,Jackiw:1976fn,Callias:1978kg},
and mention how they can be used to construct (twisted) periodic
instantons, building on the works
\cite{Garland:1988bv,Nahm:jb, 
Lee:1997vp,Kraan:1998kp,Kraan:1998pm,Kraan:1998xe,Lee:1998bb}.
\subsection{Review}
\label{self-duality} In this subsection we follow closely the line
of development in \cite{Lee:1998vu}. We first exhibit the magnetic
monopole solutions corresponding to the $su(2)$ embeddings. We
parameterize the space $R^3 \times S^1$ by $x^{1,2,3}$ and
 $x^4 \equiv x^4 + 2 \pi R$.
The self-duality equation
\begin{eqnarray}
\ast F = F
\end{eqnarray}
on $R^3 \times S^1$ can be rewritten in terms of the fields\footnote{
See also appendix \ref{app} for our conventions.} $A_i$ ($i=1,2,3$)
 and $A_4 \equiv \Phi$ as a generalized monopole equation\footnote{It is
useful to  interpret it
 as a monopole equation with structure group
the loop group \cite{Garland:1988bv}. We will postpone making use
of this interpretation to the next subsection.}:
\begin{eqnarray}
B_i = D_i \Phi - \partial_4 A_i. \label{me}
\end{eqnarray}
where $D_i = \partial_i - i [ A_i,.]$.
We choose the
Wilson line (or T-dual scalar)
 vacuum expectation value (at $x^3 = - \infty$) to be:
\begin{eqnarray}
\Phi_0 &=& v         \\
       &=& \phi_0 +  \langle v, \alpha\rangle t^3
\end{eqnarray}
where $v \in {\cal H}$, the $t_i$ form a preferred $su(2)$ subalgebra
of the gauge group (see appendix \ref{app}) and we have
\begin{eqnarray}
t^3 &=&  \frac{1}{2}\check{\alpha}
\\
\phi_0 &=& v - \frac{1}{2} \langle v, \alpha\rangle \check{\alpha}
.
\end{eqnarray}
The value for $\phi_0$ is chosen such that $[\phi_0, { t^i}]=0$.
The monopole solution corresponding to this $su(2)$ embedding
is:
\begin{eqnarray}
\label{su2sol} A_i &=& A(r,u) \epsilon_{ijk} e_r^j t^k \\ \Phi &=&
\phi_0 + \phi(r,u) e_r^i t_i
\end{eqnarray}
where we made use of the definitions
\begin{eqnarray}
A(r,u) &=& \frac{1}{r}-\frac{u}{\sinh(ur)} \\ \phi(r,u) &=&
\frac{1}{r} - u \coth(ur) \\ e_r^i &=& \frac{(x^1,x^2,x^3)}{r}
\end{eqnarray}
 and the value for $u=\langle v,\alpha\rangle$ is chosen to make sure that the
boundary conditions at infinity are satisfied.
The topological charge of the monopole is
given by:
\begin{eqnarray}
m =  \frac{1}{8 \pi ^2} \int_0^{2 \pi R} dx^4 \int_{S^2_{\infty}}
dS_i \left( B_i^a \Phi^a - \frac{1}{2} \epsilon_{ijk} A_j^a
\partial_4 A_k^a \right).
\end{eqnarray}
where we recognize a contribution from the magnetic field at
infinity, and a seemingly pure gauge part from the dependence of
the gauge field on the circle coordinate. The solutions in this
subsection did not involve the coordinate $x^4$ in a non-trivial
way, and the field $A_4=\Phi$ played the usual role of the adjoint
Higgs scalar. In this context it was shown that the solutions
along $su(2)$ embeddings which correspond to simple roots are
fundamental\footnote{The simple roots are chosen such that $v$
lies in the fundamental chamber.}
\cite{Weinberg:1980zt,Jackiw:1976fn,Callias:1978kg}.
\subsection{Extra solutions on a compact space}
\label{newsolutions}
Now we turn to the extra solutions corresponding to the fact that we
work on $R^3 \times S^1$. In section \ref{reviews} we
already pointed out that we expect one more fundamental monopole
solution to arise on this space. This is the solution we will
concentrate on.

On the compact space, we have to specify our boundary conditions.
The fields can be periodic:
\begin{eqnarray}
A_{\mu} (x^4+ 2 \pi R) = A_{\mu} (x^4).
\end{eqnarray}
or they can be periodic up to an outer automorphism $\tau$
 \begin{eqnarray}
A_{\mu} (x^4+ \frac{2 \pi R}{k}) = A_{\mu}^a(x^4) \tau (T_a),
\label{twbc}
\end{eqnarray}
where  $\tau^k=1$.
The first case was treated in detail in \cite{Lee:1998vu}. There the
extra fundamental monopole solution was found to be:
\begin{eqnarray}
\Phi(x,\theta_0) &=& \phi_0 + \frac{1}{2R}  \check{\theta_0}
+e^{-\frac{i x^4}{R } t^3(-\theta_0)} t^i(-\theta_0) e^{\frac{i
x^4}{R } t^3(-\theta_0)} e_r^i \phi(r,u_n), \nonumber  \\
A_i(x,\theta_0) &=& e^{-\frac{i x^4}{R } t^3(-\theta_0)} t^l
(-\theta_0) e^{\frac{i x^4}{R } t^3(\theta_0)} \epsilon_{ijl}
e_r^j A(r,u_n). \label{sol}
\end{eqnarray}
Here we introduced the root $\theta_0$. This root is related to
the root  $\alpha_{\epsilon}=\delta-\theta_0$, where $\delta$ is
the imaginary unit root of the affine algebra, and
$\alpha_{\epsilon}$ is the root needed to complete the simple root
system of the subalgebra $g_0$ to a simple root system of the full
affine algebra. We moreover used the definition $u_n=\langle
v,-\theta_0\rangle+\frac{1}{R}$. Now, we have chosen our
normalization in (\ref{twbc}) such that this solution also
satisfies the boundary conditions for the twisted case, and one
can check that this solution still satisfies the right boundary
conditions for the Higgs field at infinity and the generalized
monopole equation (\ref{me}). To extend (and in the end simplify)
the analysis of the properties of these solutions, it will be
useful to introduce a few more facts about affine algebras first,
then to rewrite the solutions in a more convenient and
recognizable form.

Note that for the untwisted algebras $\theta_0$ is $\theta_h$, the
highest root in $g^o$ (see footnote \ref{ghato}), and for the
twisted algebras $\theta_0$ is the highest weight of the $g_0$
representation spanned by the roots in $g_1$ \cite{K}. It will be
convenient (as in the analysis of the affine Weyl group \cite{KM})
to introduce an extended version of the vacuum expectation value
$v$. We define $v_e=v+d$, where $d$ is the derivation of the
affine algebra (normalized such that $\langle
d,\delta\rangle=\frac{1}{R}$). We moreover define the $su(2)$
subalgebra of the affine algebra corresponding to the root
$\alpha_{\epsilon}$ as follows:
\begin{eqnarray}
t^1_e(\alpha_{\epsilon}) &=& \frac{1}{2}
(E_{-\theta_0}^1+E_{\theta_0}^{-1}), \nonumber
\\
t^2_e(\alpha_{\epsilon}) &=& \frac{1}{2i}
(E_{-\theta_0}^1-E_{\theta_0}^{-1}), \nonumber
\\
t^3_e(\alpha_{\epsilon}) &=& -\frac{1}{2} \check{\theta_0}.
\label{affsu2}
\end{eqnarray}
where $E_{-\theta_0}^1=e^{-i \frac{x^4}{kR}} \otimes
E_{-\theta_0}$ (see e.g. \cite{Fuchs:1992nq}). These definitions
allow us to rewrite the solution  (\ref{sol}) in the standard form
where $\phi_0^e=v- \langle v_e,\alpha_{\epsilon}\rangle t^e_3$ and
$u_n=\langle v_e,\alpha_{\epsilon}\rangle$:
\begin{eqnarray}
\Phi(x,\alpha_{\epsilon}) &=& \phi_0^e + t^i_e(\alpha_{\epsilon})
e_r^i \phi(r,u_n), \nonumber  \\ A_i(x,\alpha_{\epsilon}) &=&
t^k_e (\alpha_{\epsilon}) \epsilon_{ijk} e_r^j A(r,u_n).
\end{eqnarray}
This simple observation will make life easier on us in the next
section, and we already understand a lot better how to interpret
the solutions found in \cite{Lee:1998vu} in a roundabout
way\footnote{The method in \cite{Lee:1998vu} can be seen as being
based on a symmetry between the zeroth root in the Dynkin diagram
of an untwisted affine algebra and another, simple root in the
Dynkin diagram.}. Indeed, the solutions correspond to the $su(2)$
subalgebra associated to the simple root $\alpha_{\epsilon}$,
embedded in the full affine algebra, much as the fundamental
monopole solutions in subsection \ref{self-duality}.
\subsection{Fundamental monopoles}
\label{properties} To prove that the monopoles constructed in the
previous section are fundamental (in the sense of
\cite{Weinberg:1980zt}), we have to analyze the number of zero
modes of the equation
\begin{eqnarray}
(-i \sigma_k D_k + D_4 ) \psi = 0
\end{eqnarray}
in the background of the monopoles. The case with periodic
boundary conditions has been treated in \cite{Lee:1998vu}. We will
simplify and extend this analysis here.  We find that the
monopoles associated to the simple roots, and with a vacuum
expectation value $v$ have four zero modes (and thus are
fundamental), when $v$ lies in the fundamental alcove of the
(twisted) affine algebra, i.e. when $ \langle v,\alpha_i\rangle$
and $\frac{1}{R}-\langle v,\theta_0\rangle$ are
positive.\footnote{We treat only the case where the symmetry group
is completely broken to $U(1)$ factors \cite{Weinberg:1980zt}.}
These conditions can be re-formulated as: $\langle v_e,
\alpha_j\rangle$ is positive for \em all \em simple roots
$\alpha_j$ of the affine algebra.

All this follows from a careful analysis of the zero mode
equation, in analogy with the analysis in appendix C of the first
reference in \cite{Weinberg:1980zt}. A  crucial observation is
that on the circle, we can expand the fluctuation mode in a basis
for the affine algebra, where the KK-momentum around the circle
correspond to the `level' of the generators (as in the definition
after equation (\ref{affsu2})). We can then define a commuting
isospin $t^3$ and hypercharge $y_{tot}$ (where $\beta$ should now
be read as taking values in the root lattice of the affine
algebra):
\begin{eqnarray}
t^3_e E_{\beta} &=& [t^3_e,E_{\beta}]=
[\frac{\check{\alpha_{\epsilon}}}{2}, E_{\beta}] \\
   &=& \frac{1}{2} \langle\check{\alpha_{\epsilon}},\beta\rangle E_{\beta}
\end{eqnarray}
and $y_{tot}$
\begin{eqnarray} y_{tot} &\equiv& \frac{1}{\langle v_e,\alpha_{\epsilon}\rangle}
[v_e-\langle v_e,\alpha_{\epsilon}\rangle t_e^3, E_{\beta}]
\nonumber
\\
  &=&
 \frac{\langle v_e,\beta\rangle}{ \langle v_e,\alpha_{\epsilon}\rangle}
-\frac{1}{2} \langle\check{\alpha_{\epsilon}},\beta\rangle.
\nonumber
\end{eqnarray}
As in \cite{Weinberg:1980zt} we can then split the
 zero mode equation into multiplets with different hypercharge
 $y_{tot}$ and isospin
quantum numbers. The rest of the proof for the fact that the monopoles we
discussed are fundamental then follows in complete parallel to the proof in
appendix C of \cite{Weinberg:1980zt}. Crucial in the proof is only the fact
that $\alpha_{\epsilon}$ is simple. This implies that the hypercharge is
larger than the isopsin ($|y|>t$) for all $\beta \neq \alpha_{\epsilon}$,
 and in that case it can be shown that
there is no normalizable solution to the zero mode equation
\cite{Weinberg:1980zt}.

Note that a periodic instanton, with zero magnetic charge and
instanton number $1$ can be formed by appropriately
\cite{Kraan:1998kp,Lee:1998bb} gathering fundamental monopoles.
For any affine algebra the multiplicity of the fundamental
monopoles in a periodic instanton is given by the Kac marks (as
pointed out for the untwisted algebras in \cite{Lee:1998vu}). The
pictorial explanation for this we already gave: this corresponds
to combining $D(p-2)$-branes until they fully wrap the circle (see
figure \ref{around}). Note that, as before, we see a one to one
correspondence between magnetic monopoles corresponding to $su(2)$
embeddings, and the co-roots of the (twisted) affine algebra.
Simple co-roots correspond to fundamental monopoles, imaginary
roots to periodic instantons, etcetera.

\section{Conclusion and discussion}

We pointed out a nice correspondence between orientifold string
theory backgrounds and (twisted) affine algebras. Apart from
clarifying string theory through algebra, it perhaps also
clarifies twisted affine algebras to string theorists. Note that
we had to introduce a new kind of orientifold action in compact
spacetimes, including an outer automorphism action on Chan-Paton
factors. This kind of projection might well generalize to more
generic compact spacetimes. A first extension of our program in
this paper might lie in the treatment of spacetimes including a
$T^2/Z_2$ factor with orientifold planes at the fixed points. The
planar positioning of the orientifold planes seems to indicate a
connection to more general Dynkin diagrams, of hyperbolic or
generalized Kac-Moody algebras, although at present we do not know
how to make this intuition precise.

Our correspondence moreover paves the way for an application in
field theory. The twisted boundary conditions that we imposed, can
be taken over in supersymmetric field theories on, for instance
$R^3 \times S^1$. One interesting recent result is the
determination of the exact elliptic superpotential for the $SU(N)$
${\cal N}=1^{\ast}$ theory on this space \cite{Dorey:1999sj}. In
that paper, a relation between general ${\cal N}=1^{\ast}$
theories on $R^3 \times S^1$ was pointed out. {From} the results
on integrable models and $N=2$ field theories summarized in
\cite{D'Hoker:1999ft}, one concludes that the relevant integrable
models \cite{Gorsky:1995zq,Martinec:1996by,Donagi:1996cf} for
general gauge groups with periodic boundary conditions are the
twisted elliptic integrable models, since they yield the right
superpotential \cite{Seiberg:1996nz,Katz:1997th,Davies:2000nw} in
the limit where one decouples the three chiral multiplets
completely.\footnote{We thank Prem Kumar for many enlightening
discussions in which this conclusion was drawn.} Our work
indicates that the other elliptic superpotentials and their
limiting affine Toda system  \cite{D'Hoker:1999ft} describe ${\cal
N}=1^{\ast}$ theories with twisted boundary conditions on the
circle, and the corresponding  ${\cal N}=1$ theories. One should
be able to check this from a first principle calculation of the
superpotential along the lines of \cite{Davies:2000nw}. It should
also be rewarding to delve more deeply into index theorems
\cite{Callias:1978kg,Bernard:1977nr,NS} on $R^3 \times S^1$ with
twisted boundary conditions to complete the zero mode analysis for
twisted periodic instantons, where, for example, we would expect
the number of zero modes of the instanton with charge one to be
given by the Coxeter number of the affine algebra (instead of the
dual Coxeter number for the usual periodic instantons).
\label{conclusion}

\section*{Acknowledgements}
We would like to thank Oliver DeWolfe, Zack Guralnik, Prem Kumar,
Joe Polchinski and Angel Uranga for discussions. This work is
supported in part by funds provided by the U.S. Department of
Energy (D.O.E.) under cooperative research agreement
DE-FC02-94ER40818. A.H. would like to thank the department of
Physics at the Weizmann Institute, the high energy theory group in
Tel Aviv University and the ITP in UCSB for their kind support
while completing various stages of this work. The research of A.H.
was supported in part by an A. P. Sloan Foundation Fellowship, by
the Reed Fund Award and by a DOE OJI Award.

\appendix
\section{Conventions}
\label{app}
The Yang-Mills field strength and magnetic field are
defined as follows:
\begin{eqnarray}
F_{\mu \nu}^a = \partial_{\mu} A_{\nu}^a
 - \partial_{\nu} A_{\mu}^a
- i [ A_{\mu}, A_{\nu} ]^a \nonumber \\ B_i = \frac{1}{2}
\epsilon_{ijk} F^{jk}.
\end{eqnarray}
For algebra conventions, we follow \cite{K,KM}, and define the set
of simple roots as
 $\Pi=\{ \alpha_1, \dots \alpha_n \} \subset {\cal H}^{\ast}$
and the set of simple co-roots as $\Pi=\{ \check{\alpha_1}, \dots,
\check{\alpha_n} \} \subset {\cal H}$. These are subsets of dual
spaces, and we have
\begin{eqnarray}
\langle\check{\alpha_i}, \alpha_i\rangle = a_{ij}
\end{eqnarray}
with $a_{ij}$ the Cartan matrix of the algebra. A special basis is given by:
\begin{eqnarray}
{[}H, E_{\alpha_i}{]} = \langle H,\alpha_i\rangle E_{\alpha_i} \\
{[} E_{\alpha}, E_{-\alpha} {]} = \check{\alpha},
\end{eqnarray}
where $H$ denotes any element of the Cartan subalgebra (including
the co-roots). We define $su(2)$ subalgebras associated to roots
$\alpha$ by:
\begin{eqnarray}
t^1(\alpha) = \frac{1}{2} (E_{\alpha}+E_{-\alpha}) \nonumber
\\ t^2(\alpha) = \frac{1}{2i} (E_{\alpha}-E_{-\alpha})
\nonumber \\ t^3(\alpha) = \frac{1}{2} \check{\alpha}.
\end{eqnarray}

\end{document}